\documentclass[aps,prl,twocolumn,superscriptaddress]{revtex4}
\usepackage{amsmath,amssymb}
\usepackage{bm}
\usepackage{graphicx}
\usepackage{epstopdf}

\newcommand{\beq}{\begin{equation}}
\newcommand{\eeq}{\end{equation}}

\begin{document}

\title{Topological disorder triggered by interaction-induced flattening of electron spectra  in solids}	
\author{V.~A.~Khodel}
\affiliation{National Research Centre Kurchatov Institute, Moscow, 123182, Russia}
\affiliation{McDonnell Center for the Space Sciences \&
Department of Physics, Washington University,
St.~Louis, MO 63130, USA}
\author{J.~W.~Clark}
\affiliation{McDonnell Center for the Space Sciences \&
Department of Physics, Washington University, St.~Louis, MO 63130, USA}
\affiliation{University of Madeira, 9020-105 Funchal, Madeira, Portugal}
\author{M.~V.~Zverev}
\affiliation{National Research Centre Kurchatov Institute, Moscow,
123182, Russia}
\affiliation{Moscow Institute of Physics and Technology, Dolgoprudny,
Moscow District 141700, Russia}
	
\begin{abstract}
We address the intervention of classical-like behavior, well documented
in experimental studies of strongly correlated electron systems of solids
that emerges at temperatures $T$ far below the Debye temperature $T_D$.
We attribute this unexpected phenomenon to spontaneous rearrangement of
the conventional Landau state beyond a critical point at which the
{\it topological} stability of this state breaks down, leading to the
formation of an {\it interaction-induced flat band} adjacent to the
nominal Fermi surface. We demonstrate that beyond the critical point,
the quasiparticle picture of such correlated Fermi systems still
holds, since the damping of single-particle excitations remains small
compared with the Fermi energy $T_F=p^2_F/2m_e$.  A Pitaevskii-style
equation for determination of the rearranged quasiparticle momentum
distribution $n_*({\bf p})$ is derived, which applies to explanation
of the linear-in-$T$ behavior of the resistivity $\rho(T)$ found
experimentally.
\end{abstract}	
		
\maketitle

Currently,``topological'' has become one of the most commonly used
terms in condensed-matter physics, surpassing ``quantum critical point.''
It is sufficient to mention such collocations as topological order,
topological transition and topological insulator.  On the other hand.
over decades the mathematical literature has featured, along with
more traditional types of chaotic behavior, relevant discussions of
topological entropy and topological chaos, which exhibit {\it positive}
entropy $S$~\cite{adler,li-yorke,blanchard,li-ye}.
(See also the Supplemental Material (SM) \cite{SM} and sources
\cite{poincare,mfeig,may,reichl,pbm,pan} cited therein)).
In the present work addressing strongly correlated electron systems of
solids including cuprates and graphene, we investigate possible existence
of a finite entropy $S>0$ at temperatures $T$ much lower than the Debye
value $T_D$ identifying the boundary between classical and quantum regimes.

Seemingly, this option would be obviated by the Nernst theorem requiring
$S(T)$ to vanish upon reaching $T=0$.  However, recent developments
warrant a revision of this conventional stance.  The first symptoms
appeared in measurements~\cite{steglich1,steglich2} of the thermal
expansion coefficient
$\alpha(T)= -V^{-1}\partial V/\partial T=V^{-1}\partial S/\partial P$
of the strongly correlated heavy-fermion superconductor CeCoIn$_5$, which
has a tiny critical value $T_c=2.3\,{\rm K}$ at which superconductivity
terminates.  Although experimental results are indeed consistent with
obedience of the Nernst theorem requiring $\alpha(0)=0$, it is
nevertheless of paramount significance that at extremely low temperatures
$T>T_c^+=T_c+0$ where the system is already in the normal state,
experiment has established the perplexing behavior
\beq
\alpha(T)=\alpha_0+\alpha_1T .
\label{nfl}
\eeq
The {\it nonzero} offset $\alpha_0\simeq 0.5\times 10^{-5}/{\rm K}$
exceeds values found in ordinary metals at these temperatures by a
huge factor of order $10^{3}-10^{4}$. This implies that an analogous
{\it classical-like} offset $S_0$, associated with $\alpha_0$ by
the relation $\alpha_0=\partial S_0/\partial P$, is present in
the entropy itself --  pointing unambiguously to the presence of
disorder in the regime of extremely low  $T>T_c^+\ll T_D$.

Another experimental challenge is associated with the low-temperature,
non-Fermi-liquid (NFL) behavior of the normal-state resistivity $\rho(T)$
of the same CeCoIn$_5$ metal at various pressures $P$, which, according
to FL theory, should obey the formula $\rho(T)=\rho_0+A_2T^2$.  Instead,
at $P<P^*\simeq$ 2\,GPa, experiment~\cite{thomp} has revealed the
{\it classical-like} strange-metal behavior
\beq
\rho(T)=\rho_0+A_1T ,
\label{rho1}
\eeq
shown in Fig.~1.
It is as if classical physics already prevails at $T_c^+<T\ll T_D$.
This remarkable linear-in-$T$ behavior of $\rho(T)$ is currently observed
in diverse systems (see e.g.~\cite{loch,gegenwart,tai1}). In some cases,
the slope $A_1$ experiences a noticeable jump~\cite{mck} (see below).

Even more bizarre behavior has surfaced in recent studies~\cite{young}
of the resistivity of twisted bilayer graphene (TBLG) as a function of
twist angle $\theta$, as depicted in Fig.~2. Profound variations of
$A_1(\theta)$ are seen, especially toward to the so-called magic angle
$\theta_m$, where the $A_1$ term increases by more than three orders of
magnitude, as does the residual resistivity $\rho_0(\theta)$, echoing
a tenfold variation of $\rho_0$ as a function of pressure $P$, as shown in
Fig.~1. Since $\rho_0$ must be a parameter-independent quantity~\cite{LL}
if the impurity population remains unchanged, its documented behavior
defies explanation within the standard FL approach.

Moreover, in high-temperature superconducting, overdoped copper oxides,
where $T_c(x)$ terminates at critical doping value $x_c$ with nearly
linear dependence on $x_c-x$ (see Fig.~3), the quite remarkable doping
independence
\beq
A_1(x)/T_c(x)={\rm const},
\label{at}
\eeq
has been discovered~\cite{hussey,bozovic}, a feature shared with the
Bechgaard salts~\cite{leyraud}.  As emphasized in Ref.~\cite{hussey},
this feature points to the presence of a hidden phase, emergent at
$x_c$ {\it simultaneously} with the superconducting state.

Explanation of the strange-metal behavior~(\ref{rho1}) observed ubiquitously
at low $T$ has become one of the most intensely debated theoretical
problems of the modern condensed-matter theory.  Analysis of proposed
scenarios in a recent review article~\cite{norman} has concluded that
none of these is capable of explanation of all the relevant experimental
findings.  In particular, candidates based on a quantum-critical-point
(QCP) scenario fall short.  As witnessed by the phase diagrams of CeCoIn$_5$,
cuprates, and graphene, there are no appropriate ordered phases adjacent
to the strange-metal region; the effects of associated quantum fluctuations
are small.

\begin{figure}[t]
\begin{center}
\includegraphics[width=0.8\linewidth] {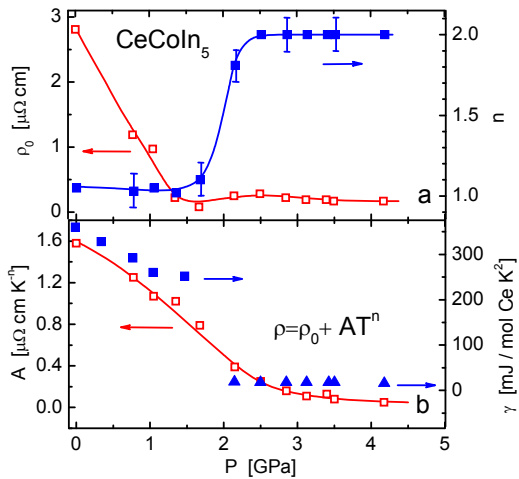}
\end{center}
\caption{Upper panel: Values of the residual resistivity $\rho_0$ (left
axis, open squares) and the index $n$ in the fit $\rho(T)=\rho_0+AT^n$
(right axis, solid squares) versus pressure $P$.  Bottom panel:
Temperature
coefficient of resistivity $A$ (left panel, open squares) and
specific-heat
coefficient $\gamma$ (right panel, solid squares and solid triangles).
The authors thank J.\ D. Thompson for permission to present data
published in Ref.~\cite{thomp}}.
\label{fig:CeCoIn5}
\end{figure}

\begin{figure}[t]
\begin{center}
\includegraphics[width=0.6\linewidth]{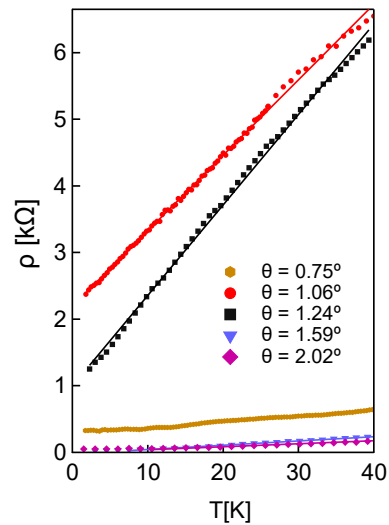}
\end{center}
\caption{NFL resistivity $\rho(T)$ measured in TBLG devices at different
twist angles. The authors express their gratitude to A.\ F.\ Young for
permission to present data published as Fig. 3i of Ref. [12] and
providing the corresponding file.}
\label{fig:TBLG}
\end{figure}

\begin{figure}[t]
\begin{center}
\includegraphics[width=0.8\linewidth] {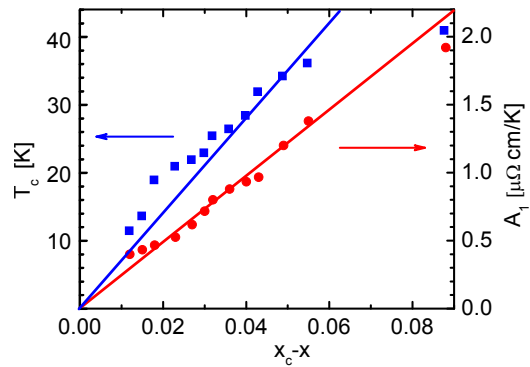}
\end{center}
\caption{Dependence of the factor $A_1$ in the resistivity $\rho(T)$
(red circles, right axis) and the critical temperature $T_c$ (blue squares,
left axis) of overdoped La$_{2-x}$Sr$_x$CuO$_4$ films on the doping $x$
measured from its critical value $x_c=0.26$ \cite{bozovic}. Red and blue
lines show the best linear fits to the data, which support the conclusion
that $A_1(x)\propto T_c(x)$, indicative of behavior inconsistent with
conventional theory.}
\label{fig:a1_tc_lsco}
\end{figure}

In this situation, we turn to a different scenario, based on the formation
of a fermion condensate
(FC)~\cite{ks,vol1,noz,physrep,vol1994,ktsn1,prb2008,annals,book}. Analogy
with a boson condensate (BC) is evident in the respective densities of
states $\rho_{\rm FC}(\varepsilon)= n_{\rm FC}\delta(\varepsilon)$~\cite{ks}
and $\rho_{\rm BC}(\varepsilon)= n_{\rm BC}\delta(\varepsilon)$, where
$n_{\rm FC}$ and $n_{\rm BC}$ are the FC and BC densities.
To be more specific, the essence of the phenomenon of fermion condensation lies
in a {\it swelling of the Fermi surface}, i.e. in emergence of an
{\it interaction-induced} flat portion $\epsilon({\bf p })=0$ of the
single-particle spectrum $\epsilon({\bf p})$ that occupies a region
${\bf p}\in\Omega$ where the real quasiparticle momentum distribution
(hereafter denoted $n_*({\bf p})$) departs drastically from the
Landau step $n_L({\bf p})=\theta(-\epsilon({\bf p}))$.

The trigger for such a profound rearrangement of the Landau state lies
in violation of its necessary stability condition (NSC), which requires
positivity of the change
$\delta E=\sum_{{\bf p}} \epsilon({\bf p})\delta n_L({\bf p})$
of the ground state energy $E$ under {\it any variation} of the $n_L({\bf p})$
compatible with the Pauli principle~\cite{physrep}.  In Landau theory with
$\epsilon(p)=v_F(p-p_F)$, this NSC is known to hold as long as the Fermi
velocity $v_F$ remains positive.  Beyond a critical point where it
breaks down, the Fermi surface becomes multi-connected.  This aspect
is a typical {\it topological} signature.  Accordingly, the word
topological in the term topological chaos has a twofold meaning, such
that the associated bifurcation point can be called a topological critical
point (TCP). Frequently, as in a neck-distortion problem addressed by
I.\ M.\ Lifshitz in his seminal article~\cite{lif}, the corresponding
topological rearrangement of the Fermi surface is unique.  However, this
is not the case in dealing with the TBLG problem, where nearly-flat-band
solutions are found, a distinctive feature  of those being related to the
{\it passage} of the Fermi velocity through zero at the first magic
twist angle $\theta^{(1)}_m$~\cite{morell,bm,neto,herrero}. A variety
of options for violation of the topological stability of the TBLG
Landau state then arise.  In contrast, within the FC scenario,
introduction of $e-e$ interactions leads to the advent of
interaction-induced flat bands which replace the
nearly-flat bands found in Refs.~\cite{morell,bm,neto,herrero}.
Technically, this procedure is reminicent of the Maxwell construction
in statistical physics, where the isotherm in the Van der Waals
pressure-volume phase diagram is in reality replaced by a horizontal
line.  An analogous situation is inherent in cuprates and other
strongly correlated electron systems of solids. Importantly, in
the familiar temperature-doping phase diagram, it is the TCP that
separates the well-understood FL behavior from the behavior associated
with topological chaos, which is responsible for the strange-metal
regime.

We begin analysis with the reminder that in superconducting alloys that obey
Abrikosov-Gor'kov theory~\cite{ag,gor}, the damping $\gamma$ acquires a
{\it finite} value due to impurity-induced scattering, implying failure of the
basic postulate $\gamma/\epsilon({\bf p})<1$ of Landau theory. Nonetheless,
the FL quasiparticle formalism, in which the pole part $G_q$ of the
single-particle Green function $G=(\epsilon-\epsilon^0_p-\Sigma)^{-1}$ has
the form
\beq
G_q({\bf p},\varepsilon)=
\frac {1-n_L({\bf p})}{\varepsilon-\epsilon({\bf p})
	+i\gamma}+ \frac {n_L({\bf p}) }{\varepsilon-\epsilon({\bf p})
	-i\gamma} ,
\label{flg}
\eeq
is still applicable~\cite{prb2019}.

Beyond the TCP where interaction-induced flat bands emerge, further
alteration of the pole part $G_q$ occurs, its form becoming
\cite{ks,vol1,noz,physrep,ktsn1}:
\beq
G_q({\bf p},\varepsilon)=
\frac {1-n_*({\bf p})}{\varepsilon-\epsilon({\bf p})
+i\gamma(\varepsilon)}+ \frac {n_*({\bf p}) }{\varepsilon-\epsilon({\bf p})
-i\gamma(\varepsilon)} ,
\label{fcqf}
\eeq
with $\gamma>0$ and occupation numbers $0<n_*({\bf p})<1$ characterizing the
FC. Their difference from $n_L({\bf p})$, which resides solely in
the $\Omega$ region, is to be determined through solution of a nonlinear
integral Landau-Pitaevskii style equation  (cf. Refs.~\cite{lan2,pit,agd})
of the theory of fermion condensation, viz.
\beq
\frac{\partial\epsilon({\bf p})}{\partial {\bf p}}=
\frac{\partial\epsilon_0({\bf p})}{\partial {\bf p}}
+2\int f({\bf p},{\bf p}_1)
\frac{\partial n_*({\bf p}_1)}{\partial {\bf p}_1}\frac{d^3{\bf p}}{(2\pi)^3} .
\label{pitc}
\eeq
Here $f({\bf p},{\bf p}_1)$ is the spin-independent part of the Landau
interaction function.  The free term includes all contributions to the
group velocity that remain in the $f=0$ limit.

A salient feature of the $T=0$ FC solutions is the {\it identical
vanishing} of the dispersion of the spectrum $\epsilon({\bf p})$ in
the $\Omega$ region. At $T>0$, the FC spectrum acquires a small
dispersion, linear in $T$~\cite{noz},
\beq
\epsilon({\bf p},T)= T\ln \frac{1-n_*({\bf p})} {n_*({\bf p})} ,
\quad  {\bf p}\in \Omega.
\label{nozs}
\eeq
Experimental verification of this effect through ARPES measurements is
crucial for substantiation of the FC concept under consideration.

Eq.~(\ref{pitc}) is derived from the formal relation $\delta \Sigma=
\bigl({\cal U} \delta G\bigr)$ (with $\delta G(p,\varepsilon)
=G({\bf p}-e{\bf A},\varepsilon)-G({\bf p},\varepsilon)$) of variational
many-body theory for the self-energy in terms of the subset of Feynman
diagrams ${\cal U}$ of the two-particle scattering amplitude that are
irreducible in the particle-hole channel, hence regular near the Fermi
surface.  Assuming gauge invariance of the theory, one finds~\cite{agd,migdal}
\beq
-\frac{\partial G^{-1}({\bf p},\varepsilon)}{\partial {\bf p}}=
\frac{\bf p}{m_e}-\biggl({\cal U} ({\bf p},\varepsilon;{\bf k},\omega)
\frac{\partial G({\bf k},\omega)}{\partial {\bf k}}\biggr).
\label{rel1}
\eeq
The round brackets in this equation imply integration
and summation over intermediate momenta and spins with a proper
normalization factor.  Implementation of a slightly refined universal
quantitative procedure [23,24] for renormalization of this equation
allows it, irrespective of correlations, to be recast in closed form,
as if one were dealing with a {\it gas of interacting quasiparticles}.
(The word ``gas'' is appropriate, since Eq.~(\ref{pitc}) contains only
the single phenomenological amplitude $f$ of quasiparticle-pair
collisions).  A salient feature of this procedure is that Eq.~(\ref{pitc})
holds {\it both} in conventional Fermi liquids {\it and} in electron
systems of solids moving in the periodic external field of the crystal
lattice.  This follows because {\it solely} gauge invariance was assumed
in its derivation, which therefore holds for crystal structures as well.
In short, the widespread impression that the FL approach is inapplicable
to crystal structures is groundless.  We emphasize once more that the
FL renormalization procedure works properly {\it irrespective
of the magnitude of the ratio} $\gamma/\epsilon({\bf p})$
(see the SM \cite{SM} for specifics, and especially references
\cite{lan1,lan2,pit,prb2019,ag}).

We are now in position to consider the connection between the customary
iterative procedure for solving the basic FC equation~(\ref{pitc}) and
the topological chaos problem addressed in many mathematical articles (see
especially \cite{blanchard,li-ye}). In the standard iterative scheme,
the $j$th iteration $n^{(j)}({\bf p})$, with $j=0,1,2,...$, is inserted
into the right side of Eq.~(\ref{pitc}) to generate the next iteration
of the single-particle spectrum, $\epsilon^{(j+1)}({\bf p})$, and this
process is repeated indefinitely to finally yield a convergent result
whose topological entropy (TE) is equal to 0.  However, beyond the
TCP, such a procedure fails, since the iterations $n^{(j)}({\bf p})$
then undergo chaotic jumps from 0 to 1 and vice-versa, generating
noise, identified with some TE. To evaluate the spectrum quantitatively, in
Ref.~\cite{prb2008}, the iterative discrete-time map was reconstructed
in such a way that the discrete time $t_j$ replaces the iteration
number $j$.  Subsequent time-averaging of relevant quantities, adapted
from formulas of classical theory, allows one to find a specific
self-consistent solution. Its prominent feature is the development of
an interaction-induced flat portion in the single-particle spectrum
$\epsilon({\bf p})$ that embraces the nominal Fermi surface
(for exemplification, see the SM \cite{SM}).
Another distinctive signature of the set of specific FC solutions of
Eq.~(\ref{pitc}) lies in the occurrence of a nonzero entropy excess $S_*$,
emergent upon their substitution into the familiar combinatoric formula for
evaluation of the entropy.  This yields~\cite{ks,prb2008,annals,prb2012}
\beq
S_*= - 2\sum_{\bf p}  [n_*({\bf p}) \ln n_*({\bf p})+(1-n_*( {\bf p}))
\ln (1-n_*({\bf p}))] ,
\label{en}
\eeq
where summation is running over the FC region,
and, in turn, a NFL nonzero value $\alpha_*$ of the coefficient of thermal
expansion.  Because the presence of a nonzero $S_*$ would contradict the
Nernst theorem $S(T=0)=0$ if it survived to $T=0$, the FC must inevitably
disappear~\cite{ks,annals,prb2012} at some very low $T$. One well-elaborated
scenario for this metamorphosis is associated with the occurrence of phase
transitions, such as the BCS superconducting transition emergent in the
case of attraction forces in the Cooper channel, or an antiferromagnetic
transition, typically replacing the superconducting phase in external
magnetic fields $H$ exceeding the critical field $H_{c2}$.

At $H<H_{c2}$, a nonzero BCS gap $\Delta(0)$ in the single-particle spectrum
$E({\bf p})=\sqrt{\epsilon^2({\bf p})+\Delta^2} $ does provide for
nullification of $S(T=0)$.  This scenario applies in systems that host a
FC as well, opening a specific route to high-$T_c$ superconductivity
\cite{ks,khv,vol3}.  Indeed, consider the BCS equation for determining
$T_c$:
\beq
D({\bf p}) =-2\int{\cal V}({\bf p},{\bf p}_1)
\frac{\tanh\frac{\epsilon({\bf p}_1,T_c)}{2T_c}}
{2\epsilon({\bf p}_1,T_c)}D({\bf p}_1)dv_1.	
\label{tc}
\eeq
Here $D({\bf p})= \Delta({\bf p},T\to T_c)/\sqrt{T_c-T}$ plays the role
of an eigenfunction of this linear integral equation, while
${\cal V}({\bf p},{\bf p}_1)$ is the block of Feynman diagrams for the
two-particle scattering amplitude that are irreducible in the Cooper
channel.  Upon insertion of Eq.~(\ref{nozs}) into this equation and
straightforward momentum integration over the FC region, one arrives at
a non-BCS {\it linear} relation
\beq
T_c(x)=c(x)\,T_F,
\label{tfc}
\eeq
where $c(x)=\lambda\,\eta(x)$, with $\lambda$ denoting the effective
pairing constant and $\eta(x)$ the FC density. This behavior is in accord
with the experimental Uemura plot~\cite{uemura,herrero}.

As discussed above, the entropy excess $S_*\propto \eta$ comes into play
at temperatures $T^+_c< T\ll T_D$ so as to invoke a $T$-independent term
$\alpha_0$ in the coefficient of thermal expansion, which, in that regime,
serves as a signature of fermion condensation~\cite{annals,prb2012}.
Accordingly, execution of extensive low-$T$ measurements of the thermal
expansion coefficients in candidate materials would, in principle, provide
means (i) to distinguish between flat bands that do not entail
excess entropy $S_*$ and the interaction-induced exemplars, and (ii) to
create a database of systems that exhibit pronounced NFL properties,
in aid of searches for new  exotic superconductors.

Very recently, the FC scenario has gained tentative support from ARPES
measurements performed in monolayer graphene intercalated by Gd, which
have revealed the presence of a flat portion in the single-particle
spectrum~\cite{link}.  However, verification of the correspondence
between the flat bands detected in the bilayer system
TBLG~\cite{morell,bm,herrero, yankovitz,herrero2,herrero3,young} and
the interaction-driven variety considered here requires a concerted
analysis of kinetic properties, especially of comprehensive
experimental data on the low $T$ resistivity $\rho(T)=\rho_0+A_1T+A_2T^2$.

Numerous theoretical studies of the NFL behavior of $\rho(T)$ based on
the FC concept have been performed.  Directing the reader to
Refs.~\cite{prb2012,prb2013,pla2018} for details, we summarize
their pertinent results in the relation
\beq
\rho_0(x,P,\theta)=\rho_i+a_0\eta^2(x,P,\theta), \, A_1(x,P,\theta)
=a_1\eta(x,P,\theta),
\label{rhot}
\eeq
where $\rho_i$ is the impurity-induced part of $\rho$ and $a_0,a_1$ are
factors independent of input parameters.  This expression properly explains
the data shown in Figs.~1 and 2.  Indeed, we see that in systems having
a FC, the residual resistivity $\rho_0$ depends critically on the FC
density $\eta$, which changes under variation of input parameters such
as doping $x$, pressure $P$, and twist angle $\theta$ -- an effect that
is missing in the overwhelming majority of extant scenarios for the
NFL behavior of the resistivity $\rho(T)$.  Comparison of Eq.~(\ref{rhot})
with Eq.~(\ref{tfc}) shows that the theoretical ratio $A_1(x)/T_c(x)$ is
indeed doping-independent, in agreement with the challenging experimental
results shown in Fig.~3.  Moreover, assuming that the FC parameter
$\eta(x)$ varies linearly with $x_c-x$, which is compatible with model
numerical calculations based on Eq.~(\ref{pitc}), the corresponding
result obtained from Eq.~(\ref{rhot}) is consistent with available
experimental data~\cite{pla2018}.

Turning to the issue of classical-like Planck dissipation~\cite{zaanen},
we observe that such a feature is inherent in systems that possess a
specific collective mode, transverse zero-sound (TZS), which enters
provided $m^*/m_e > 6$~\cite{halat}.  (Notably, in LSCO this ratio exceeds
10~\cite{taillefer}, while in CeCoIn$_5$ it is of order $10^2$~\cite{mac}).
In the common case where the Fermi surface is multi-connected, some
branches of the TZS mode turn out to be damped, thereby ensuring the
occurrence of a linear-in-$T$ term in the resistivity $\rho(T)$. This
is broadly analogous to the situation that arises for electron-phonon
scattering in solids in the classical limit $T>T_D$. (See also
the SM \cite{SM} and Refs.~\cite{jetpl2010,mig_100,mac,halat}.)
As a result, FC theory predicts that a break
will occur in the straight line $\rho(T)= \rho_0+A_1T$ at some
characteristic Debye temperature $ T_{\rm TZS}$~\cite{prb2013,shag2019},
in agreement with experimental data on Sr$_3$Ru$_2$O$_7$~\cite{mac}.
However, in the case $T_{\rm TZS}<T_c$, often inherent in exotic
superconductors such as CeCoIn$_5$~\cite{thomp} and TBLG
(M4 device)~\cite{herrero3}, this break disappears, and the behavior
of $\rho(T)$ is fully reminiscent of that in classical physics.

In contrast, a current scenario of Patel and Sachdev (PS) for Planckian
dissipation~\cite{sachdev}
attributes the NFL behavior (2) of $\rho(T)$ to the presence of
a significant random component in the amplitude of the interaction
between quasiparticles near the Fermi surface.  However, such a mechanism
is hardly relevant to the physics of cuprates.  Indeed, in their phase
diagrams, the strange-metal regions located above respective high-$T_c$
domains are commonly adjacent to the familiar FL ones, whose properties
obey Landau FL theory, in which the interaction amplitudes are free from
random components. Hence the PS model can be a toy model at best.
Otherwise, boundaries of the high-$T_c$ regions must {\it simultaneously}
be points of phase transitions between FL phases and phases with random
behavior of the interaction amplitudes, which is unlikely.

In conclusion, we have demonstrated that the concept of topological
chaos is capable of explaining the non-Fermi-liquid, classical-like
behavior of strongly correlated electron systems that is emergent at
temperatures $T$ far below the Debye value $T_D$, where such behavior
hitherto seemed impossible. The origin of the topological chaos,
especially well pronounced in graphene, is shown to be associated
with the presence of interaction-induced flat bands.  The theoretical
predictions are  consistent with  experimental findings, as
documented in Figs.~1-3.

We are grateful to P.\ Esquinazi, P.\ Gegenwart, M.\ Greven, E.\
Henriksen, M.\ Katsnelson, Ya.\ Kopelevich, S.\ Kravchenko, Z.\ Nussinov, F.\ Steglich,
V.\ Shaginyan, J.\ D.\ Thompson, and G.\ Volovik for fruitful
discussions. VAK and JWC acknowledge financial support from the
McDonnell Center for the Space Sciences.

\newpage

\renewcommand{\theequation}{S\arabic{equation}}
\setcounter{equation}{0}

\renewcommand{\thefigure}{S\arabic{figure}}
\setcounter{figure}{0}

\begin{widetext}

\centerline{\bf SUPPLEMENTAL MATERIAL}
\vskip .5cm

\begin{center}
{\bf Topological disorder triggered by interaction-induced flattening of electron spectra  in solids}	
\end{center}

\begin{center}
{V.~A.~Khodel, J.~W.~Clark, M.~V.~Zverev}
\end{center}

\vskip 0.5cm

\end{widetext}

\vskip 1.cm

\parindent 0pt
{\bf 1. Topological entropy and fermion condensation: an elemental
example}
\vskip .4cm

\parindent 11pt
The relationship of fermion condensation to the mathematical concept
of topological entropy \cite{blanchard,adler,li-yorke,li-ye}
and operations performed in treating classical chaos
\cite{poincare,mfeig,may,reichl} can be demonstrated in the
following example from high-energy physics.

Superdense quark-gluon plasma (QGP) is unique among quantum many-body
systems in that the topological rearrangement of the Landau state leading
to a fermion condensate already occurs in first-order perturbation
theory, which yields a schematic single-particle spectrum
\beq
\epsilon(p)-\mu =v_F(p-p_F)+a(p-p_F)\ln (p-p_F) .
\label{pmb1}
\eeq
This result is reminiscent of the single-particle spectrum of a homogeneous
Coulomb plasma, with a crucial difference: the QGP parameter $a$
carries a {\it positive} sign, due to the attractive character of
quark-gluon exchange, leading to an infinite {\it negative} slope of the
spectrum $\epsilon(p)$ at the Fermi momentum $p=p_F$.  As demonstrated
in the iterative procedure applied by Pethick, Baym, and Monien (PBM)
\cite{pbm}, this difference has phenomenal consequences, giving
rise to unlimited breeding of new sheets of the Fermi surface.

The same PBM Fermi-sheet proliferation persists in a more accurate
evaluation of the QGP spectrum based on the Dyson equation
\cite{prb2008}
\beq
\epsilon(p)=\epsilon^0_p+g\int \ln\frac{1}{|p-p_1|} n(p_1)dp_1,
\label{deqgp}
\eeq
where $\epsilon^0_p\simeq cp$ is the bare spectrum with $c$ the velocity
of light, $g>0$ is an effective coupling constant, and $n(p)$
is the corresponding quasiparticle momentum distribution.

Since an attempt at straightforward iterative solution of Eq.~(\ref{deqgp})
fails, an alternative approach was applied in Ref.~\cite{prb2008}.
As outlined below, the problem was reformulated in terms of an iterative
discrete-time map, in analogy with basic treatments of classica dynamical
chaos (e.g., \cite{mfeig}), with subsequent time-averaging of relevant
quantities.  The salient feature of the resulting self-consistent solution
of Eq.~(\ref{deqgp}) is the development of an interaction-induced flat
portion in the single-particle spectrum $\epsilon(p)$, embracing the
nominal Fermi surface.

\begin{figure}[t]
\begin{center}
\includegraphics[width=1.0\linewidth]{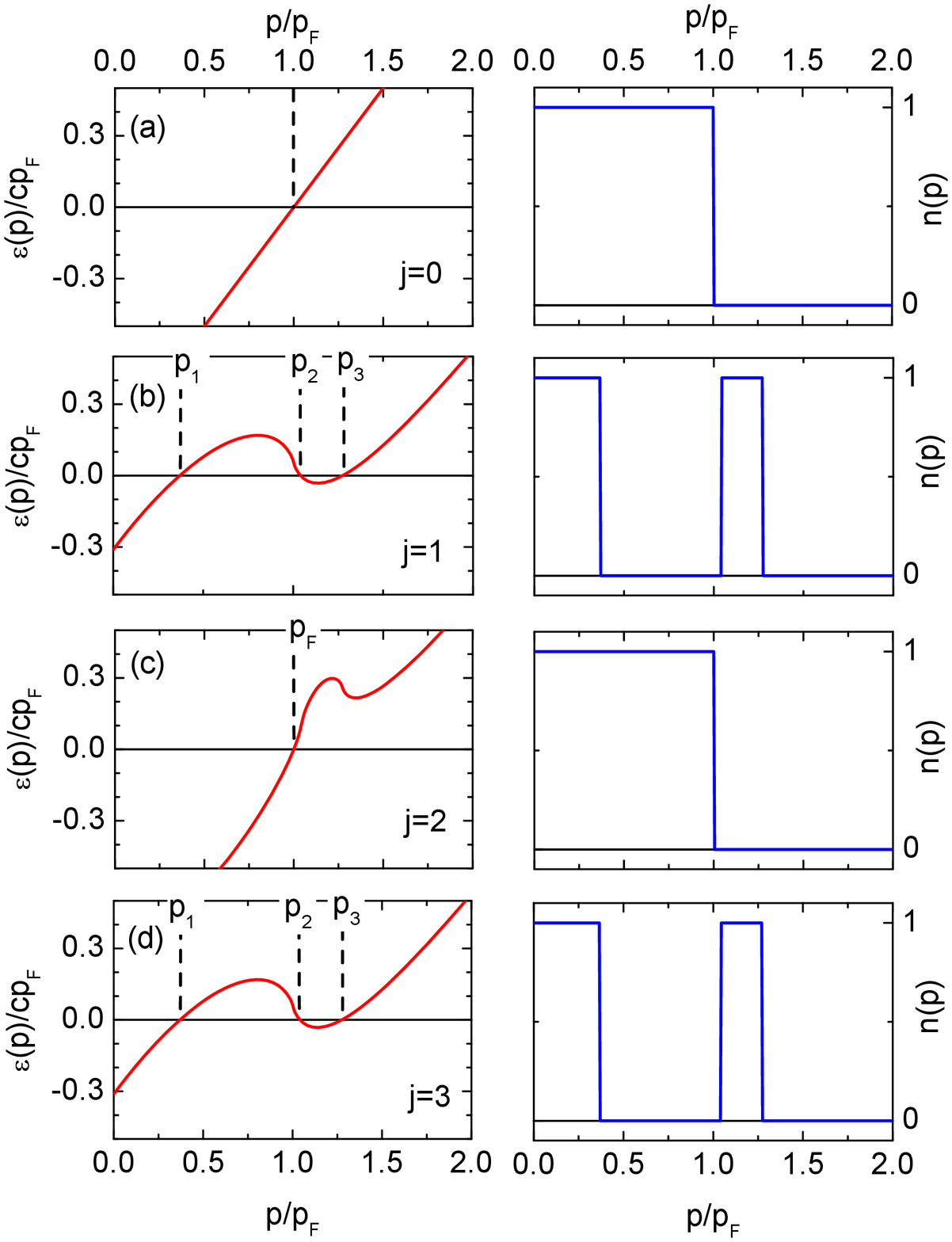}
\end{center}
\caption{Iterative maps for Eq.~(\ref{deqgp}) with the dimensionless
parameter $g/c$ set to unity. The left-hand panels show iterations
for the spectrum $\epsilon^{(j)}(p)$ in units of $cp_F$ for $j=0,1,2,3$
(these iterations being reckoned from the corresponding iterations for
the chemical potential), while the right-hand panels show iterations
for the momentum distribution $n^{(j)}(p)$.}
\label{fig:qgp_2c_c}
\end{figure}

In the standard iterative scheme, the $j$th iteration $n^{(j)}(p)$, with
$j=0,1,2...$, is inserted into the right side of Eq.~(\ref{deqgp}) to
generate the next iteration of the single-particle spectrum,
$\epsilon^{(j+1)}(p)$, measured now from the chemical potential $\mu$,
and this process is repeated indefinitely.  As seen in
Fig.~\ref{fig:qgp_2c_c},
the divergence of the slope of $\epsilon^{(0)}(p)$ leads originally to
a specific 2-cycle in which all even iterations coincide with the Landau
momentum distribution $n_L(p)=\theta(p_F-p)$.  Coincidence occurs in
all odd iterations as well; however, their structure differs from that
of the Landau state by the presence of a break in $n_{FL}(p)$.  In principle,
this cycle is eliminated in a more sophisticated iterative scheme in which
the neighboring iterations are mixed with each other~\cite{prb2008,pan}.
However, such a refinement does not lead to convergence.

\begin{figure}[t]
\begin{center}
\includegraphics[width=0.6\linewidth] {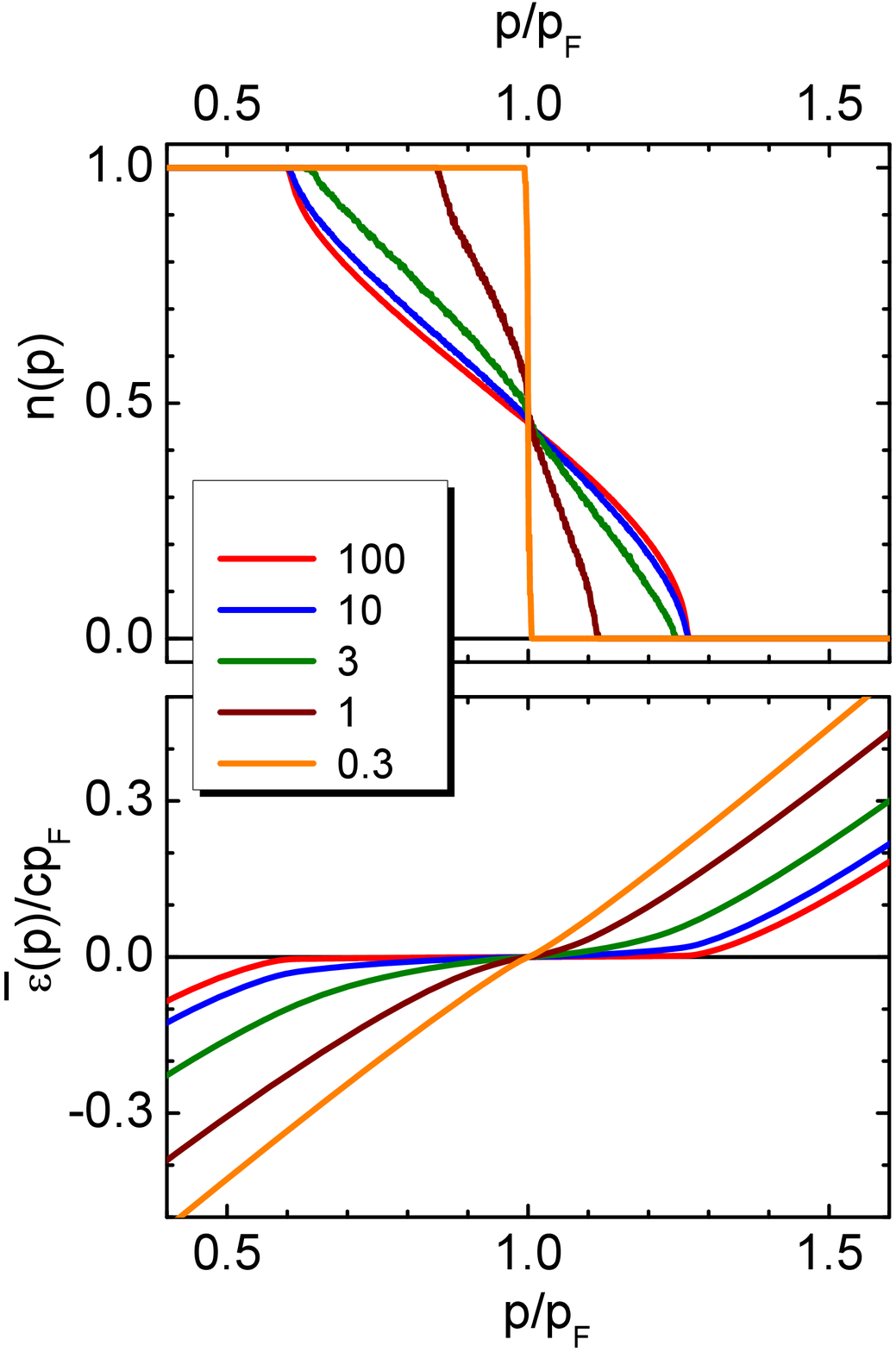}
\end{center}
\caption{Single-particle spectrum (upper panel) and momentum distribution
(lower panel) averaged according to Eq.~(\ref{enav}).  Correspondence between
the color of lines and the number of effective iterations $N\zeta$, the
product of the number of real iterations $N$, and the parameter of mixing
of neighboring iterations $\zeta$, is indicated.
}
\label{fig:qgp_av_mix_c}
\end{figure}

To overcome this difficulty, it is beneficial to introduce discrete
time-steps numbered $j$ at which the functions $\epsilon(p,t) $ and
$n(p,t)$ are updated, the latter undergoing chaotic jumps from 0 to 1
and vise-versa as $j$ and hence $t$ increases. A further adjustment makes
the crucial difference.  Aided by formulas adapted from classical mechanics,
$\epsilon(p,t)$ and $n(p,t)$ are replaced, respectively, by an averaged
single-particle spectrum $\overline{\epsilon}(p)$ and a corresponding
averaged occupation number $ \overline{n}(p)$:
\begin{eqnarray}
\overline{\epsilon}(p)\!\!&=&\!\!\!\lim_{\tau\to\infty}\frac{1}{\tau}\!
\int\limits_0^\tau
\epsilon(p,t)dt=\lim_{N\to\infty}\frac{1}{N}\sum\limits_0^N\epsilon^{(j)}(p),
\nonumber\\
\overline{n}(p)\!\!&=&\!\!\!\lim_{\tau\to\infty}\frac{1}{\tau}\!
\int\limits_0^\tau n(p,t)dt=\lim_{N\to\infty} \frac{1}{N}
\sum\limits_0^N n^{(j)}(p).
\label{enav}
\end{eqnarray}
Importantly, Eq.~(\ref{deqgp}) holds if one inserts the time-averaged quantities
$\overline{\epsilon}(p)$ and ${\overline{n}}(p)$ instead of the original ones.

Results from calculations demonstrating emergence of the interaction-induced
flat bands of the QGP are displayed in Fig.~\ref{fig:qgp_av_mix_c}. As seen,
the function $\epsilon(p)$ does vanish identically in a momentum region $\Omega$
where the 2-cycle originally sets in, while the quasiparticle momentum
distribution, denoted by $n_*(p)$, emerges as a {\it continuous} function of
$p$ in this region.
\vskip .6cm

\parindent 0pt
{\bf 2. Equality of quasiparticle and particle numbers in strongly correlated
Fermi systems}
\vskip .4cm

\parindent 11pt
In this section we outline a quasiparticle formalism free of the Landau
restriction $\gamma/\epsilon(p)\ll 1$, where $\gamma$ represents the damping
of single-particle excitations and $\epsilon$ their energy.  In strongly
correlated electron systems such as graphene, the Landau quasiparticle
picture is putatively inapplicable, since the ratio $\gamma/\epsilon$
is known not to be small, in contrast to requirements of the original
Landau theory~\cite{lan1,lan2,pit}. However, we shall find
that smallness of this ratio is a sufficient, but not necessary, condition
for validity of the quasiparticle pattern. Indeed, in superconducting
alloys, the quasiparticle formalism is operative
(cf.\ the textbook~\cite{agd} and/or Ref.~\cite{prb2019}), despite
the fact that $\gamma/\epsilon$ greatly exceeds unity due to the presence,
in $\gamma$, of a {\it  finite} term arising from energy-independent
impurity-induced scattering.  Moreover, the actual requirement for
validity of the quasiparticle method hinges on the smallness of the
damping $\gamma$ compared with the {\it Fermi energy} $T_F$, which is
met in the overwhelming majority of electron systems of interest.
Consequently,
in correlated homogeneous Fermi liquids, the particle number always coincides
with the quasiparticle number, {\it irrespective} of the magnitude
of the damping of single-particle excitations, as long as $\gamma\ll T_F$.

Proof of this statement is based on the dichotomy characterizing the
impact of long-wave external fields $V({\bf k}\to 0,\omega\to 0)$ on
correlated Fermi systems, which depends crucially on the ratio $\omega/k$.
Indeed, due to the fictitious character of coordinate-independent
external fields $V(k=0,\omega)$, no physical change of the system occurs
upon their imposition. On the contrary, change does ensue in the complementary
case of static fields $V(k,\omega=0)$, its principal effect being expressed
in the pole parts $G_q$ of the Green functions $G$.

By way of illustration, in what follows we adopt a pole part of the form
\beq
G_q(p,\varepsilon)= \bigl(\varepsilon-\epsilon(p)
+i\gamma{\rm sgn}\,(\varepsilon)\bigr)^{-1} ,
\label{lqp}
\eeq
with $\gamma>0$, as introduced by Abrikosov and Gor'kov in their theory of
superconducting alloys~\cite{ag}. However, the final results are
invariant with respect to the explicit form of $G_q$.
	
Further, in the ensuing analysis it is instructive to represent the
quasiparticle density $n$ as an integral
\beq
n= - 2\int\!\!\!\!\int p_n\frac{\partial G_q(p,\varepsilon)}
{\partial p_n} \frac{d^3{\bf p}\,d\varepsilon}{(2\pi)^4i} ,
\label{qnr}
\eeq
where $p_n$ is the momentum component normal to the Fermi surface.
Here, integration over energy is assumed to be performed before
differentiation with respect to momentum $p$.  The correct result
is also obtained provided the derivative $\partial G_q/\partial p_n$
in the integrand of Eq.~(\ref{qnr}) is rewritten in the form
\beq
\frac{\partial G_q(p,\varepsilon)}{\partial p_n} =
- \lim_{{\bf k}\to 0}G_q({\bf p},\varepsilon)\,G_q({\bf p}+{\bf k},\varepsilon)
\frac{\partial  G^{-1}_q( p, \varepsilon)}{\partial p_n},
\eeq
yielding
\beq
n= 2 \int\!\!\int\!\! p_n \lim_{{\bf k}\to 0}G_q({\bf p},\varepsilon)
\,G_q({\bf p}+{\bf k},\varepsilon)\frac{\partial G^{-1}_q(p,\varepsilon)}
{\partial p_n} \frac{ d^3{\bf p}\,d\varepsilon}{(2\pi)^4i} .
\label{qnr1}
\eeq
Evidently, integration over energy in Eq.~(\ref{qnr1}) produces a nonzero
result only if the poles of $G({\bf p},\varepsilon)$ and
$G({\bf p}+{\bf k},\varepsilon)$ lie on opposite sides of the energy axis.
This requirement is met provided the energies $\epsilon(p)$ and
$\epsilon({\bf p}+{\bf k})$ have opposite signs, so as to generate the
relevant factor $(dn(p)/d\epsilon(p))(d\epsilon(p)/dp_n)\equiv dn(p)/dp_n$
in the integration over energy.  The analogous relation
\beq
\rho= 2 \int\!\!\int\!\! p_n \lim_{{\bf k}\to 0}G({\bf p},\varepsilon)
\,G({\bf p}+{\bf k},\varepsilon) \frac{\partial G^{-1}(p,\varepsilon)}
{\partial p_n} \frac{ d^3{\bf p}\,d\varepsilon}{(2\pi)^4i} \label{pnr}
\eeq
applies for the total density $\rho$.

Hereafter we adopt symbolic notations frequently employed in Fermi Liquid
(FL) theory. With round brackets implying summation and integration over
all intermediate variables and the normalization factor $1/(2\pi)^4i$,
Eqs.~(\ref{qnr}) and (\ref{pnr}) then become
\beq
n= \biggl(p_n G_qG_q\frac{\partial G^{-1}_q}{\partial p_n}\biggr),
\quad \rho= \biggl(p_n GG\frac{\partial G^{-1}}{\partial p_n}\biggr).
\label{sn}
\eeq

Following Pitaevskii \cite{pit}, we exploit two generic identities of many-body theory.  The first of these,
\beq
- \frac{\partial G^{-1}(p,\varepsilon)}{\partial p_n} = \frac{p_n}{m} +
\biggl({\cal U}(p,k) \frac{\partial G(k,\omega)}{\partial k_n}\biggr) ,
\label{rel_1}
\eeq
where ${\cal U}$ represents the block of Feynman diagrams for the
scattering amplitude that are irreducible in the particle-hole channel,
is derived assuming gauge invariance of the theory~\cite{agd}.
The second, of the form
\beq
\frac{\partial G^{-1}(p,\varepsilon)}{\partial \varepsilon}p_n=p_n +
\biggl({\cal U}(p,k)\frac{\partial G(k,\varepsilon)}{\partial\varepsilon}k_n
\biggr),
\label{rel_2}
\eeq
stems from the commutativity of the momentum operator with the total
Hamiltonian of the system~\cite{migdal}.

The first step in the proof of the equality $\rho=n$ relies on Landau's
decomposition of the product of two single-particle Green functions into
a sum of terms,
\beq
\lim_{{\bf k}\to 0}G({\bf p},\varepsilon)\,G({\bf p}+{\bf k},\varepsilon)
 =z^2\,A(p,\varepsilon)+B( p,\varepsilon) ,
\label{deco}
\eeq
in which $B$ is a part of the limit regular near the Fermi surface,
while the remaining pole part is a product of the quasiparticle propagator
\beq
A(p,\varepsilon)= \lim_{{\bf k}\to 0}G_q({\bf p},\varepsilon)\,
G_q({\bf p}+{\bf k},\varepsilon) ,
\label{Alim}
\eeq
and a square of the quasiparticle weight
$z=\left(1-\partial\Sigma(p,\varepsilon)/\partial\varepsilon\right)^{-1}$
in the single-particle state.

It should be emphasized that there exists an important
formula~\cite{lan2,agd} analogous to Eq.~(\ref{Alim}), namely
$$
\frac{\partial G(p,\varepsilon)}{\partial \varepsilon}
= -\lim_{\omega\to 0}G( p,\varepsilon)\, G(p,\varepsilon+\omega)
\frac{\partial  G^{-1}(p,\varepsilon)}{\partial\varepsilon }
$$
\vskip -0.5 cm
\beq
\equiv-B(p,\varepsilon)\frac{\partial  G^{-1}( p,\varepsilon)}
{\partial\varepsilon}.
\eeq
Evidently, the result of integration of this expression over energy
vanishes identically, since poles of the product $G( p,\varepsilon)\,
G(p,\varepsilon+\omega)$ lie on the same side of the energy axis.
Further, one finds that regular contributions to the key relations
involved come from both the regular parts $B$ of the product $GG$ and
the block ${\cal U}$ itself.

The key step in the regularization procedure developed by Landau then
lies in the introduction of a specific interaction amplitude
$\Gamma^\omega$ determined by the equation
\beq
\Gamma^\omega={\cal U}+ \Bigl({\cal U}B\Gamma^\omega\Bigr)\equiv {\cal U}
+\Bigl(\Gamma^\omega B{\cal U}\Bigr),
\label{gamo}
\eeq
which is capable of absorbing all the regular contributions,
{\it irrespective} of the explicit form of the propagator $A$.  Indeed,
let us multiply both members of Eq.~(\ref{rel_1}) from the left
by the product $\Gamma^\omega B $, integrate over all variables,
and eliminate the expression $\Gamma^\omega B{\cal U}$ in the
last term with the aid of Eq.~(\ref{gamo}), yielding finally
$$
-\biggl(\Gamma^\omega(p,k)B(k,\varepsilon)\frac{\partial G^{-1}(k,\varepsilon)}
{\partial k_n}\biggr)= \biggl(\Gamma^\omega(p,k)B(k,\varepsilon)\frac{k_n}{m}
\biggr)
$$
\vskip -0.5 cm
\beq
+\biggl(\Gamma^\omega(p,k)\frac{\partial G(k,\varepsilon)}{\partial k_n}\biggr)
- \biggl({\cal U}(p,k)\frac{\partial G(k,\omega)}{\partial k_n}\biggr) .
\eeq

Simple algebraic transformations then lead to the relation
\beq
-\frac{\partial G^{-1}(p,\varepsilon)}{\partial p_n}
=\frac{\partial G^{-1}(p,\varepsilon)}{\partial
\varepsilon} \frac{p_n}{m} -\biggl(z^2\Gamma^\omega A
\frac{\partial G^{-1}(k,\omega)}{\partial k_n}\biggr).
\label{pitfc}
\eeq
In obtaining this result, we have employed the
equation~\cite{pit,agd,migdal}
\beq
\frac{\partial G^{-1}(p,\varepsilon)}{\partial\varepsilon}p_n=p_n
 +\Bigl(\Gamma^\omega Bk_n\Bigr) .
\label{pit1}
\eeq

Near the Fermi surface, the relevant derivatives of $G^{-1}$ are
evaluated in terms of the corresponding derivatives of the pole
part $G^{-1}_q$ to obtain
\beq
-\frac{\partial G^{-1}_q(p,\varepsilon)}{\partial p_n}
=\frac{\partial G^{-1}_q(p,\varepsilon)}{\partial \varepsilon}\frac{p_n}{m}+
\biggl( z^2\Gamma^{\omega}\frac{\partial G_q(k,\omega)}{\partial k_n}\biggr).
\label{pit2}
\eeq
Significantly, integration over energy is obviated due to the presence
of the Fermi surface, providing the structure of the pole term $G_q$
is given by Eq.~(\ref{lqp}).

Thus, in homogeneous matter, we are led to a Pitaevskii-style equation
\beq
\frac {\partial\epsilon(p)}{\partial p_n}=\frac{p_n}{m}+
\biggl(f\frac{\partial n_*(k)}{\partial k_n}\biggr) ,
\eeq
where the Landau notation $f= z^2\Gamma^\omega $ is introduced. We
emphasize that the sole condition employed has been the smallness
of the ratio $\gamma/T_F\ll 1$.

Beyond the critical point where the topological stability of the Landau
state breaks down, the momentum distribution $n_*(p)$ acquires the
non-Fermi-Liquid (NFL) form found from numerical solution of this
equation.  With regard to TBLG and similar electron systems, the
first term on the right side of this equation must be improved,
as has been done for example in Ref.~\cite{bm}.

To prove the coincidence $\rho=n$, we multiply both sides of
Eq.~(\ref{pitfc}) from the left by the product $p_nB$ and integrate
over all intermediate variables.  Exploiting the fact that the integral
$\Bigl(B\,\partial G^{-1}(p,\varepsilon)/\partial\varepsilon\Bigr)\equiv
-\Bigl(\partial G(p,\varepsilon)/\partial\varepsilon\Bigr)$
vanishes identically in integrating over the energy, while observing
that the remaining term involving the product $p_nB\Gamma^\omega$
simplifies with the aid of relation (\ref{pit1}), we are led to
$$
\biggl(p_n B \frac{\partial G^{-1}(p,\varepsilon)}{\partial p_n}\biggr) =
\biggl(p_nBz^2\Gamma^\omega A\frac{\partial G^{-1}(k,\varepsilon)}
{\partial k_n}\biggr)
$$
\vskip -0.5 cm
$$
=z^2\biggl(p_n A\frac{\partial G^{-1}(p,\varepsilon)}{\partial\varepsilon}
\frac{\partial G^{-1}(p,\varepsilon)}{\partial p_n}
\biggr)
$$
\vskip -0.5 cm
\beq
- z^2\biggl(p_n A\frac{\partial G^{-1}(p,\varepsilon)}
{\partial p_n}\biggr) .
\label{equa}
\eeq
Transfer of the last term in Eq.~(\ref{equa}) to the left side of this
equation and further straightforward manipulation of the relation
obtained leads us finally to
\beq
\rho=\frac{2}{3}\biggr(p_nG_q(p,\varepsilon)\,G_q(p,\varepsilon)
\frac{\partial G_q^{-1}(p,\varepsilon)}{\partial p_n}
\biggr)=n ,
\eeq
thereby establishing the coincidence of particle and quasiparticle numbers,
independently of the magnitude of the damping of single-particle
excitations.  In the end, all that matters is the presence of the
Fermi surface.

The final result $\rho=n$ is not sensitive to the explicit form of the
quasiparticle propagator $A$. Consequently, it can be understood that the
equality between particle and quasiparticle numbers generalizes to Fermi
systems having Cooper pairing with a BCS gap function $\Delta({\bf p})$.
In that case it takes the form
\beq
\rho=n=2\sum_{\bf p} v^2({\bf p}) \equiv
2\sum_{\bf p}\biggl( \frac{1}{2}-\frac {\epsilon({\bf p})}
{\sqrt{\epsilon^2({\bf p})+\Delta^2({\bf p}})}
\biggr)
\label{pqc}
\eeq
irrespective of the structure of the single-particle spectrum, thereby
allowing for the presence of flat bands and/or damping of normal-state
single-particle excitations.  This statement, best proved within the
framework of the Nambu formalism along the same lines as before,
provides the hitherto missing element of the FL approach to the theory
of strongly correlated superconducting Fermi systems.
\vskip 4cm

\parindent 0pt
{\bf 3. New branches of the collective spectrum and low-$T$ kinetic
properties of strongly correlated electron systems}
\vskip .4cm

\parindent 11pt
Here we discuss consequences stemming from the occurrence of additional
branches of collective excitations in strongly correlated electron systems
of solids~\cite{jetpl2010,mig_100} whose presence is associated with
a substantial enhancement of the effective mass $m^*$ -- often exceeding
$10^2m_e$ as in heavy fermion metals, notably CeCoIn$_5$~\cite{mac}.
Primarily, one is dealing with the transverse zero sound (TZS) emergent when the
first dimensionless harmonic of the Landau interaction function satisfies
$F_1=f_1p_Fm^*/\pi^2>6$~\cite{halat}.

We briefly outline results from Ref.~\cite{mig_100}, in which a standard
assumption for the arrangement of the Fermi surface is adopted.  It
is assumed to consist of two sheets, one containing light carriers
whose Fermi velocity $v_L\simeq p_F/m_e$ greatly exceeds that of the
heavy carriers populating the second band, $v_H=p_F/m^*$. In such
systems, there exist several branches of TZS.  Here
we focus on a mode whose velocity is smaller than the Fermi value $v_L$.
In this case, the dispersion relation yielding the complex value of its
velocity $c=c_R+ic_I$ takes the form~\cite{mig_100}
$$
1=\frac{F_1}{6}\biggl[1-3\biggl(\frac{c^2}{v^2_H}-1\biggr)\biggl(\frac{c}{2v_H}
\ln\frac{c+v_H}{c-v_H}-1\biggr)\biggr] $$
\beq
+\frac{F_1v_H}{6v_L}\biggl[1-3\biggl(\frac{c^2}{v^2_L}-1\biggr)
\biggl(\frac{c}{2v_L} \ln\frac{c+v_L}{ c-v_L}-1\biggr)\biggr] .
\label{eq1}
\eeq
The imaginary part of the expression on the right side of this equation
vanishes identically.  As is easily verified, its real part comes
primarily from the first term in the square brackets on the right-hand
side, since the second term is suppressed by presence of the small
factor $v_H/v_L$.  In the realistic case $F_1\gg 1$, we arrive after
some algebra~\cite{mig_100} at the result
\beq
c_R\propto v_L\sqrt{\frac{m_e}{m^*}}, \quad c_I\propto v_L\frac{m_e}{m^*}.
\eeq
Thus we see that in strongly correlated electron systems, there is no ban
on emission and absorption of the TZS quanta, by virtue
of the condition $c_R/v_L< 1 $. One is then allowed to study the associated
collision term along the same lines as in the familiar case of
electron-phonon interactions in solids, where the resistivity $\rho(T)$
varies linearly with $T$ provided $T>T_D$.  Additionally, the velocity
$c_R$ typically turns out to be smaller than the phonon velocity.  This
becomes the essential factor, especially in the case where the relevant
Debye temperature becomes lower than $T_c$ and kinetic properties of
the normal states then obey classic laws in the whole temperature region.

\end{document}